\documentclass{raa}            % referee version: for submission

\usepackage{graphicx,times}             %for PS/EPS graphics inclusion, new

\begin{document}

  \title{On the variable timing behavior of PSR B0540$-$69: an almost excellent example to study pulsar braking mechanism
%\footnotetext{$*$ Supported by the National Natural Science Foundation of China.}
}
 % \subtitle{On the variable timing behavior of pulsars}

   \volnopage{Vol.0 (200x) No.0, 000--000}      %%preserved for Editor. DOn't remove!
   \setcounter{page}{1}          %%starting page, preserved for Editor. DOn't remove!

     \author{F. F. Kou\inst{1,2}
     \and Z. W. Ou\inst{1,2}
   \and H. Tong
      \inst{1}
   }

   \institute{Xinjiang Astronomical Observatory, Chinese Academy of Sciences, Urumqi 830011,
    China; {\it tonghao@xao.ac.cn}\\
    \and University of Chinese Academy of Sciences, 19A Yuquan Road, Beijing, China
           }

   \date{Received~~2015 month day; accepted~~2015~~month day}

\abstract{PSR B0540$-$69 has braking index measurement in its
persistent state: $n=2.129 \pm 0.012$. Recently, it is reported to
have spin-down state changes: a suddenly $36\%$ increase in the
spin-down rate. Combining the persistent state braking index
measurement and different spin-down states, PSR B0540$-$69 is more
powerful than intermittent pulsars in constraining pulsar
spin-down models. The pulsar wind model is applied to explain the
variable timing behavior of PSR B0540$-$69. The persistent state
braking index of PSR B0540$-$69 is the combined effect of magnetic
dipole radiation and particle wind. The particle density reflects
the magnetospheric activity in real-time and may be responsible
for the changing spin-down behavior. Corresponding to the $36\%$
increase in the spin-down rate of PSR B0540$-$69, the relative
increase in the particle density is $88\%$ in the vacuum gap
model. And the model calculated braking index in the new state is
$n=1.79$. Future braking index observation of PSR B0540$-$69 in
the new spin-down state will be very powerful in distinguishing
between different pulsar spin-down models and different particle
acceleration models in the wind braking scenario. The variable
timing behavior of PSR J1846$-$0258 is also understandable in the
pulsar wind model.
 \keywords{pulsars: general -- pulsars: individual (PSR B0540$-$69; PSR J1846$-$0258) -- stars: neutron -- wind} }

   \authorrunning{Kou, Ou \& Tong}            %author_head in even pages
   \titlerunning{On the variable timing behavior of pulsar B0540$-$69}  % title_head in odd pages

   \maketitle
%% The author head (on even pages) and the title head (on odd pages) will be
%% automatically extracted from \author{} and \title{}. Whenever the title is too long,
%% you will be asked to supply a shorter one by inserting either \authorrunning{} or
%% \titlerunning{} before \maketitle. Anyway, you can specify your own heads.
%%
%%
%% Note: In the following text body of your manuscript, please note several differences from
%%       other major journals:
%% (1) \subsection{Please Capitalize the First Letter of Each Notional Word in Subsection Title}
%% (2) Please Capitalize the First Letter of Each Notional Word in all tables' captions

%
%________________________________________________ sections below
%
\section{Introduction}           %% first-level sections will be auto-capitalized

PSR B0540$-$69, known as the ``Crab Twin'', is a young radio
pulsar with spin-down parameters $\nu \approx 19.727 \rm \,Hz$,
$\dot{\nu} \approx -1.86 \times 10^{-10} \rm \, Hz  s^{-1}$
(Marshall et al. 2015) and braking index $n=2.129\pm0.012$
(Ferdman et al. 2015). Its characteristic magnetic field is about
$10^{13} \, \rm G$ at the magnetic poles\footnote{Assuming all the
rotational energy is consumed by magneto-dipole radiation in
vacuum, $B({\rm pole})=6.4 \times 10^{19} \sqrt{P \dot{P}} \,\rm
G$}. Only two glitches with relative small changes in spin-down
parameters were reported (Zhang et al. 2001; Cusumano et al. 2003;
Livingstone et al. 2005; Ferdman et al. 2015). Recently, a
persistent and unprecedented increase in the spin-down rate of PSR
B0540$-$69 was observed: the relative increase in the spin-down
rate is $36\%$ which is orders magnitude larger than the changes
induced by glitches (Marshall et al. 2015). Another pulsar PSR
J1846$-$0258 was also reported to have variable timing behaviors:
a net decrease in the spin frequency ($\Delta\nu \approx
-10^{4}\rm \, Hz$) after the large glitch (Livingstone et al.
2010) and a lower braking index $n=2.19\pm 0.03$ (Livingstone et
al. 2011; Archibald et al. 2015b) than its persistent state value
$n=2.65\pm 0.01$ (Livingstone et al. 2006).

 The spin-down behavior of pulsars can be described by the power law:
 \begin{equation}
 \label{nudotpowerlaw}
 \dot{\nu}=- C \nu^{n},
 \end{equation}
 where $\nu$ and $\dot{\nu}$ are respectively the spin frequency and frequency derivative,
 $C$ is usually taken as a constant and $n$ is the braking index.
 The braking index is defined accordingly:
 \begin{equation}
 \label{defn}
 n=\frac{\nu \ddot{\nu}}{\dot{\nu}^2},
 \end{equation}
where $\ddot{\nu}$ is the second derivative of spin frequency. The
braking index reflects the pulsar braking mechanism (Tong 2015).
In the magneto-dipole braking model, a pulsar rotates uniformly in
vacuum $ \dot{\nu}\propto \nu^{3}$. The expected braking index is
three which is not consistent with the observations (Lyne et al.
2015). Like the intermittent pulsar (Kramer et al. 2006), PSR
B0540$-$69 also has two different spin-down states. For the
intermittent pulsar PSR B1931$+$24, people tried to measure its
braking index during the on and off state (Young et al. 2013). Now
this aim has been partially fulfilled by PSR B0540$-$69 which has
not only different spin down states but also braking index
measurement for the persistent state (``low'' spin-down rate
state), see Table \ref{parameters}. Therefore, it can put more
constraints on pulsar spin-down models. Any candidate model should
explain both the braking index during the persistent state and the
variable spin-down rate.

Previously, the pulsar wind model (Xu \& Qiao 2001) is employed to
explain the spin-down behavior of intermittent pulsars (Li et al.
2014) and braking index of the Crab pulsar (Kou \& Tong 2015). In
the following, it is shown that both the persistent state braking
index and varying spin-down rate of PSR B0540$-$69 are
understandable in the wind braking model. The varying spin-down
rate is due to a variable particle wind. And the varying braking
index of PSR J1846$-$0258 is caused by a changing particle
density. The pulsar wind model and calculations are listed in
Section 2. Discussions and conclusions are presented in Section 3
and Section 4, respectively.

\begin{table}
\label{parameters}
\begin{center}
\caption{Comparison of spin-down parameters of PSR B1931$+$24, PSR
B0540$-$69, and PSR J1846$-$0258. The intermittent pulsar PSR
B1931$+$24 has different spin-down states without any braking
index information at present. PSR B0540$-$69 has both different
spin-down states and the persistent state braking index
measurement. PSR J1846$-$0258 is reported to have a variation of
braking index.}

\begin{tabular}{llllll}
\hline \hline
Pulsar name & $\nu (\rm Hz)$ & $\dot{\nu} (\rm Hz \, s^{-1})$ & braking index\\
\hline
B1931$+$24 $(\rm off)^{a}$ & $1.229$ & $-10.8 \times 10^{-15}$ & $?$ \\

B1931$+$24 $(\rm on)^{a}$ & $1.229$ & $-16.3 \times 10^{-15}$ &
$?$ \\\hline

B0540$-$69 $(\rm low)^{b}$ & $19.727$ & $-1.86 \times 10^{-10}$ & $2.129^{\rm c}$ \\

B0540$-$69 $(\rm high)^{b}$ & $19.701$ & $-2.53 \times 10^{-10}$ &
$?$ \\\hline

J1846$-$0258 (persistent state)$\rm ^d$ & 3.08 & $-6.72\times 10^{-11}$ & 2.65 \\

J1846$-$0258 (after glitch)$\rm ^e$ & 3.06 & $-6.65 \times 10^{-11}$ & 2.19 \\

 \hline
\end{tabular}
\flushleft (a): From Kramer et al. (2006). The on state has larger spin-down rate than the off state. \\
(b): From  Marshall et al. (2015). ``Low'' means the previous
spin-down state and ``high'' means the new spin-down state
with a higher spin-down rate.\\
(c) Mean value of braking index (Ferdman et al. 2015).\\
(d) From Livingstone et al. (2006). \\
(e) From Archibald et al. (2015b).
\end{center}
\end{table}

\section{Variable timing behavior of pulsars caused by a varying particle wind}

\subsection{Description of the pulsar wind model}

Pulsars are oblique rotators in general. The perpendicular and
parallel magnetic dipole moment may respectively relate to the
magnetic dipole radiation and particles acceleration (Xu \& Qiao
2001; Kou \& Tong 2015):
\begin{equation}
\dot{E_{\rm d}}=\frac{2 \mu^{2} \Omega^{4}}{3 c^{3}}
\sin^{2}\alpha , \label{Edotdipole}
\end{equation}

\begin{equation}
\dot{E_{\rm p}}=2 \pi r_{\rm p}^{2} c \rho_{\rm e} \Delta \phi
=\frac{2 \mu^{2} \Omega^{4}}{3c^{3}} 3 \kappa \frac{\Delta
\phi}{\Delta \Phi} \cos^2 \alpha, \label{Edotp}
\end{equation}
where  $\mu=1/2 B R^{3}$ is the magnetic dipole moment ($B$ is the
polar magnetic field and $R$ is the neutron star radius), $c$ is
the speed of light, and $\alpha$ is the angle between the
rotational axis and the magnetic axis (i. e., inclination angle),
$\Omega=2 \pi \nu$ is the angular velocity of the pulsar, $r_{\rm
p}=R(R \Omega/c)^{1/2}$ is the polar cap radius, $\rho_{\rm
e}=\kappa \rho_{\rm GJ}$ is the primary particle density where
$\rho_{\rm GJ}=\Omega B/(2 \pi c)$ is the Goldreich-Julian charge
density (Goldreich \& Julian 1969) and $\kappa$ is the
dimensionless particle density, $\Delta \phi$ is the corresponding
acceleration potential of the acceleration region, and $\Delta
\Phi=\mu \Omega^{2}/c^{2}$ is the maximum acceleration potential
for a rotating dipole (Ruderman \& Sutherland 1975). The pulsar
rotational energy is consumed by the combined effect of magnetic
dipole radiation and particle acceleration (Xu \& Qiao 2001)
\begin{equation}
-I \Omega \dot \Omega = \frac{2 \mu^{2} \Omega^{4}}{3 c^{3}} \eta,
\label{Edot}
\end{equation}
where $I=10^{45} \, \rm {g \, cm^{2}}$ is the moment of inertia,
and
\begin{equation}\label{equationeta}
\eta = \sin^2 \alpha + 3 \kappa \Delta \phi/{\Delta \Phi} \cos^2
\alpha.
\end{equation}
The spin-down behavior can be expressed as:
\begin{equation}
\label{Omegadotwind} \dot{\Omega}=-\frac{2 \mu^{2} \Omega^{3}}{3 I
c^3} \eta.
\end{equation}

According the equation (\ref{defn}), the braking index in the
pulsar wind model can be written as (Xu \& Qiao 2001):
\begin{equation}
n = 3 + \frac{\Omega}{\eta} \frac{d \eta}{d \Omega}  ,
\label{nwind}
\end{equation}
The exact expression of $\eta$ (equation (\ref{equationeta}))
depends on the particle acceleration potential. The vacuum gap
model (Ruderman \& Sutherland 1975) is taken as an example to show
the calculation process and
\begin{equation}\label{etaVGCR}
\eta=\sin^2\alpha+4.96 \times 10^{2} \kappa B_{12}^{-8/7}
\Omega^{-15/7} \cos^2 \alpha,
\end{equation}
where $B_{12}$ is the magnetic field in units of $10^{12} \, \rm
G$ (Kou \& Tong 2015). For other acceleration models, the
corresponding expressions of $\eta$  are listed in Table 2 in Kou
\& Tong (2015).

\subsection{On the variable timing behavior of PSR B0540$-$69}

A generic picture for the variable timing behavior of PSR
B0540$-$69 and PSR J1846$-$0258 is: a glitch may occurred during
the observations, as that in PSR J1846$-$0258 (Livingstone et al.
2010). This small glitch may have been missed in the case of PSR
B0540$-$69. This glitch may induce some magnetospheric activities,
e.g., outburst (Gavriil et al. 2008). The particle outflow will be
stronger during this process. This will cause the pulsar to have a
larger spin-down rate (Marshall et al. 2015). After some time, a
larger spin-down state will result in a net spin-down of the
pulsar compared with previous timing solutions (Livingstone et al.
2010). The braking index will be smaller since the particle wind
is stronger (Wang et al. 2012a). When the pulsar magnetosphere
relax to its persistent state, if the particle density is still
varying with time $\kappa=\kappa(t)$, the braking index will be
different with the persistent state while the change in spin-down
rate may be negelected (Livingstone et al. 2011; Archibald et al.
2015b; Kou \& Tong 2015). From previous observations of PSR
J1846$-$0258, its pulse profile has no significant variations
before, during and after the outburst (Livingstone et al. 2010,
2011; Archibald et al. 2015b). The observations of magnetar 1E
1048.1$-$5937 showed that the pulsed flux may not be a good
indicator of the magnetospheric activities (Archibald et al.
2015a). The variation of total X-ray flux is needed. Therefore,
the enhanced spin-down rate in PSR B0540$-$69 without changes in
pulse profile and pulsed flux is not unusual (Marshall et al.
2015). The reason may be that the geometry of the pulsar is
unchanged during the magnetospheric activities. This may result in
a constant pulse profile and pulsed flux.

For PSR B0540$-$69, giving the persistent state spin-down
parameters $\nu=19.727 \, \rm Hz$, $\dot{\nu}=-1.86 \times
10^{-10} \, \rm Hz \,s^{-1}$ (Marshall et al. 2015), and
inclination angle $\alpha=50^\circ$ (the best fitted value given
by Zhang \& Cheng 2000). Parameters of magnetic field $B=10^{13}
\, \rm G$ and $\kappa=834$ can be calculated (by solving equations
(\ref{Edot}) and (\ref{nwind})) corresponding to the observed
braking index $n=2.129 \pm 0.012$ (Ferdman et al. 2015). The
calculated $\kappa=834$ means that the particle density is $834$
times the Goldreich-Julian charge density which is consistent with
previous conclusions (Kou \& Tong 2015 and references therein).

\begin{figure}
\centering
\includegraphics[width=0.65\textwidth]{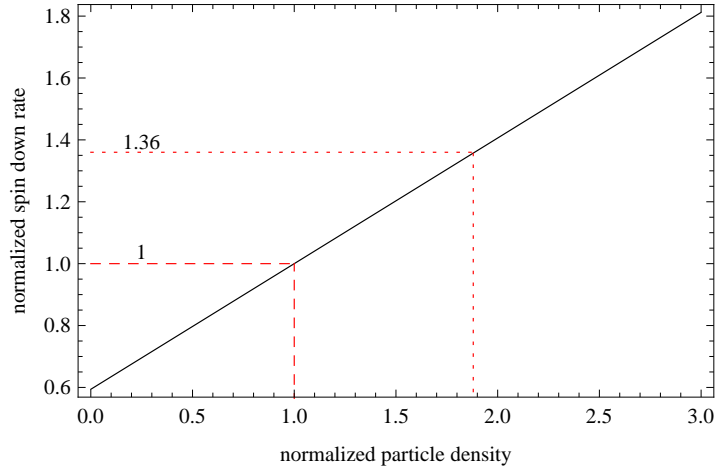}
\caption{The normalized spin-down rate as function of the
normalized particle density for PSR B0540$-$69 in the vacuum gap
model. The dashed line is the spin-down rate in the persistent
state. The dotted line is the new spin-down rate which is 1.36
times the persistent state spin-down rate (Marshall et al. 2015).}
\label{fignudotk}
\end{figure}

\begin{figure}
\centering
\includegraphics[width=0.65\textwidth]{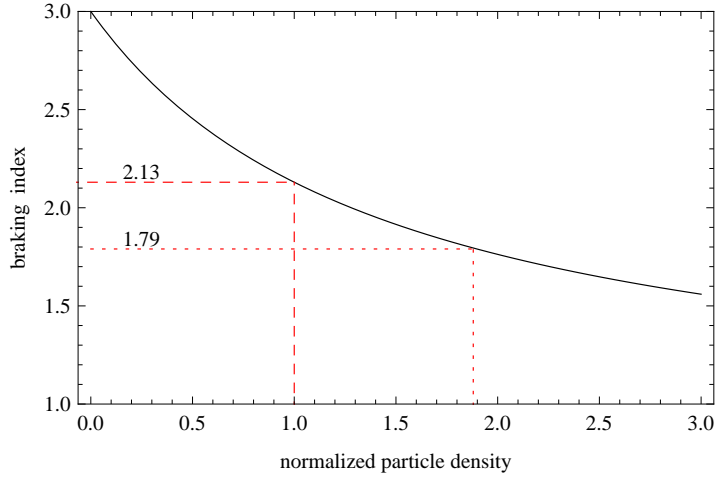}
\caption{Braking index as function of the normalized particle
density for PSR B0540$-$69 in the vacuum gap model. The dashed
line is the persistent state braking index $2.13$ (Ferdman et al.
2015). The dotted line is the braking index $1.79$ which is
predicted by the increased particle density.} \label{fignk}
\end{figure}

The spin-down rate of PSR B0540$-$69 has increased by $36\%$ in
the new spin-down state (Marshall et al. 2015). In the pulsar wind
model, the variation of the spin-down rate is caused by a
different particle density
\begin{equation}
\label{nudotwind}
\frac{\dot{\Omega}^{'}}{\dot\Omega}=\frac{\eta(\kappa^{'})}{\eta(\kappa)},
\end{equation}
where $\dot{\Omega}^{'}$ and $\eta({\kappa^{'}})$ correspond to
the new spin-down state. A larger particle density will result in
a higher spin-down rate (equation(\ref{Omegadotwind}) and
(\ref{etaVGCR})). Figure \ref{fignudotk} and figure \ref{fignk}
shows respectively the normalized spin-down rate
$\dot{\nu}^{'}/\dot{\nu}$ and braking index as function of
normalized particle density $\kappa^{'}/\kappa$ for PSR B0540$-$69
in the vacuum gap model. As shown in figure \ref{fignudotk}, the
spin-down rate increases as the particle density increases. An
increase in the particle density of $88 \%$ will result in the $36
\%$ increase in the spin-down rate. As particle density increase,
the braking index will decrease because the effect of  particle
wind component is increasing (figure \ref{fignk}). When the
particle density increases to $1.88$ times the previous value, the
braking index decreases to $1.79$, the relative change is
$15.7\%$. And the corresponding frequency second derivative will
be $\ddot{\nu}=5.83 \times 10^{-21} \, \rm Hz \,s^{-2}$ (equation
($\ref{defn}$)).

Calculations in all the acceleration models are also made. The
same conclusion is obtained from these models: an increasing
particle density results in the increase in spin-down rate. For
PSR B0540$-$69, corresponding to the observational $36\%$ relative
increase in spin-down rate, the relative increase in the particle
density in all these models ranges from $72\%$ to $154\%$. The
second frequency derivative ranges from $4.5\times 10^{-21} \, \rm
Hz\, s^{-2}$ to $6.15\times 10^{-21}\,\rm Hz \,s^{-2}$. Braking
indices in the new state in all these acceleration models are list
in Table \ref{brakingindices}. If the conversion efficiency of
particle energy to X-ray luminosity is unchanged (Becker 2009),
the total X-ray luminosity may also have increased by the same
factor.

\begin{table}

\begin{center}

\caption{Braking indices of PSR B0540$-$69 in the new state in all
the acceleration models.} \label{brakingindices}

{\scriptsize
 \begin{tabular}{llllllllll}
 \hline \hline

Acceleration models & VG(CR) & VG(ICS) & SCLF(II,CR) & SCLF(I) &
OG &
CAP & NTVG(CR) & NTVG(ICS)\\
\hline Braking index & $1.79$ & $1.87$ & $1.90$ & $1.79$ & $1.38$
&
$1.83$ & $1.90$ & $1.86$ \\
\hline
\end{tabular}}

 \flushleft
Notes: See Table 2 of Kou \& Tong 2015 for the meanings of the
acceleration models abbreviations. The minimum braking index of
the SCLF (II ICS) is $2.4$ (Li et al. 2014) which is larger than
the persistent braking index of PSR B0540$-$69: $n=2.129$. It
means that the SCLF (II, ICS) model can be ruled out or can not
exist alone to accelerate particle in the magnetosphere of PSR
B0540$-$69 (Wu et al. 2003; Li et al. 2014).
\end{center}

\end{table}

\subsection{On the variable timing behaviors of PSR J1846$-$0258}

Spin-down parameters and persistent state braking index of PSR
J1846$-$0258 are respectively: $\nu=3.08 \,\rm Hz$ and
$\dot{\nu}=-6.72 \times 10^{-11} \,\rm Hz \, s^{-1}$ and
$n=2.65\pm 0.01$ (Table\ref{parameters}) (Livingstone et al.
2006). In the pulsar wind model, corresponding to the
observational braking index, the magnetic field $B=1.25\times
10^{14} \, \rm G$ and particle density $\kappa=28$ are calculated
in the vacuum gap model with an inclination angle $45^{\circ}$ (a
inclination angle of $45^{\circ}$ is chosen in the following
calculations\footnote{There is no observational or best fitted
inclination angle given.}). Such a magnetic field is comparable
with the characteristic magnetic field $9.7\times 10^{13} \, \rm
G$ at the poles and much larger than magnetic fields of normal
pulsars. Then it is not surprising that magnetar activities can be
observed in this source (Gavriil et al. 2008).

Variable timing behavior of a net decrease in the spin-down
frequency ($\Delta \nu \approx -10^{-4}\,\rm Hz$) was detected for
PSR J1846$-$0258 after a larger glitch (Livingstone et al. 2010).
The correspondingly relative increase in the spin-down rate is
about $7\%$ ($\Delta\dot{\nu}=-4.82 \times 10^{-12} \,\rm Hz
\,s^{-1}$ during an epoch $240$ days when phase coherency is
lost). Such an increase in spin down rate may also be caused by a
larger particle density. Just like the calculation for PSR
B0540$-$69, in the vacuum gap model of the pulsar wind model, a
$44\%$ increase in the particle density results in the $7 \%$
increase in the spin-down rate. The braking index will also be
smaller during this enhanced spin-down epoch. However, braking
index measurement is only available long after the glitch when the
timing noise is greatly reduced.

\begin{figure}[!ht]
\centering
\includegraphics[width=0.65\textwidth]{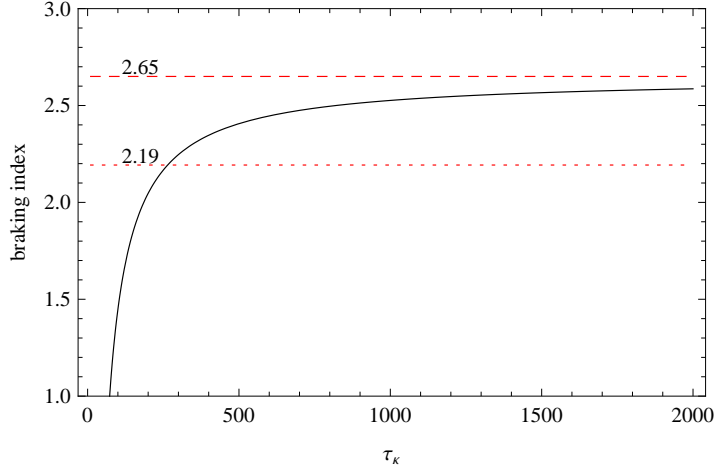}
\caption{The braking index of the PSR J1846$-$0258 as function of
$\tau_{\kappa}$ in the vacuum gap model. The dashed line is the
persistent state braking index $2.65$ (Livingstone et al. 2006).
The dotted line is the smaller braking index $2.19$ measured after
the glitch (Archibald et al. 2015b).} \label{figntauk}
\end{figure}

A lower braking index $2.19\pm 0.03$ is detected after the glitch
(Livingstone et al. 2011; Archibald et al. 2015b) which is
significantly smaller than its persistent state value $n=2.65\pm
0.01$ (Livingstone et al. 2006). In the pulsar wind model, it can
be understood by a time varying particle density
$\kappa=\kappa(t)$ (similar to the Crab pulsar, Kou \& Tong 2015):
\begin{equation}
n=3 + \frac{\Omega}{\eta} \frac{d \eta}{d \Omega}-
\frac{\kappa}{\eta} \frac{d \eta}{d \kappa}
\frac{\tau_{c}}{\tau_{\kappa}}, \label{ntauk}
\end{equation}
where $\tau_{c}=- \frac{\Omega}{2 \dot{\Omega}}$ is the
characteristic age, $\tau_{\kappa}=\frac{\kappa}{2 \dot \kappa}$
is the typical variation timescale of the particle density. An
increasing particle density ($\tau_{\kappa}>0$ or
$\dot{\kappa}>0$) lead to a smaller braking index. From figure
\ref{figntauk}, the braking index is insensitive to
$\tau_{\kappa}$ when it is much larger that $1000$ yr, but
decrease sharply when $\tau_{\kappa}$ is comparable with the
characteristic age (about $700$ yr). The changing rate of the
particle density $\dot{\kappa}=1.68\times 10^{-9} \, \rm s^{-1}$
will result in a braking index of $2.19$. During the epoch from
MJD 55369 to MJD 56651, the particle density has increased
$0.66\%$. The relative increase in spin-down rate is $0.1\%$ which
is very small. Therefore, the changing particle density will
mainly result in a different braking index while not affecting the
spin-down rate. This is the difference between the variable timing
behavior of PSR 0540$-$69 and PSR J1846$-$0258.

\section{Discussions}

Observations of intermittent pulsars (Kramer et al. 2006) and
measurement of braking indices (Lyne et al. 2015) help to
distinguish between different pulsar spin-down mechanisms. The
variable spin-down rate of PSR B0540$-$69 combined with its
persistent state braking index measurement is more powerful than
intermittent pulsars in constraining different models. In the
magneto-dipole radiation model $\dot{\nu} \propto \mu^2 \sin^2
\alpha /I \times \nu^3$, in order to explain the braking indices
($n<3$) of eight young pulsars, an increasing inclination angle
(Lyne et al. 2013), increasing magnetic field (Espinoza et al.
2011) or decreasing moment of inertia (Yue et al. 2007) is
expected. The variable spin-down rate may be induced by the change
of inclination angle, magnetic field or moment of inertia.
Corresponding to the $36\%$ relative increase in the spin-down
rate, the relative change should be: $26\%$ increase in
inclination angle, or $17\%$ increase in magnetic field, or $26\%$
decrease in moment of inertia. It seems impossible to achieve such
huge changes during a short timescale (about 14 days, Marshall et
al. 2015). A change in inclination angle is unlikely since the
pulse profile did not change significantly (Marshall et al. 2015).
An increase in magnetic field will require an increase of magnetic
energy by $36\%$, about $10^{42} \,\rm erg$. It is unlikely there
is such a huge amount of energy injection. A decrease of moment of
inertia will require a decrease of neutron star radius. During
this process, a huge amount of gravitational energy will be
released (Zhou et al. 2014), about $10^{52} \,\rm erg$, which is
again unlikely.

Previous models for the spin-down behavior of intermittent pulsars
(Beskin \& Nokhrina 2007; Li et al. 2012) may also be applied to
the variable spin-down rate of PSR B0540$-$69. However, the
expected braking index is three in Beskin \& Nokhrina (2007) and
the magnetohydrodynamical simulations (Li et al. 2012).
Considering the effect of pulsar death or evolution of inclination
angle, the braking index will be larger than three (Contopoulos \&
Spitkovsky 2006; Philippov et al. 2014). Therefore, these models
should be modified before they can explain both the persistent
state braking index and variable spin-down rate of PSR B0540$-$69.

There are several models designed for magnetar spin-down which may
also be employed to the case of PSR B0540$-$69. The magnetar
spin-down may be dominated by a particle wind (Harding et al.
1999). The calculations in Harding et al. (1999) is equivalent to
assuming each outflow particle can attain the maximum acceleration
potential of a rotating dipole (Tong et al. 2013). This wind
braking model of magnetars was employed by Kramer et al. (2006) to
explain the spin-down behavior of the first intermittent pulsar
PSR B1931$+$24. An additional particle outflow in the on state
will result in a larger spin-down rate. The rotational energy loss
is related to the particle wind luminosity $L_{\rm p}$ as $\propto
\sqrt{L_{\rm p}}$ (Harding et al. 1999). A particle wind
luminosity $85\%$ larger will result in a spin-down rate $36\%$
larger. The particle wind luminosity is related to the polar cap
radius $R_{\rm pc}$ and magnetospheric opening radius $r_{\rm
open}$ as $L_{\rm p} \propto R_{\rm pc}^4 \propto r_{\rm
open}^{-2}$ (Harding et al. 1999). Therefore, the magnetospheric
opening radius will be $26\%$ smaller. However, there are several
problems when applying the wind braking model of magnetars to the
case of normal pulsars: (1) In the wind braking model of
magnetars, a strong particle wind is assumed. The effect of
magnetic dipole radiation is neglected. This may be appliable to
the case of magnetars whose emissions are dominated by magnetic
energy output (Tong et al. 2013). However, in the case of normal
pulsars (including intermittent pulsars) the effect of dipole
radiation may not be neglected. (2) In the case of strong particle
wind, the braking index is $n=1$ (Tong et al. 2013). This is not
consistent with the braking index of pulsars (Lyne et al. 2015).
(3) When applying to the case of intermittent pulsars (Kramer et
al. 2006; Young et al. 2013), pure magnetic dipole braking is
assumed for the off state. This may be valid for the case of
intermittent pulsars whose radio emissions are stopped in the off
state (Li et al. 2014). However, this assumption can not be
applied to the persistent spin-down state of PSR B0540$-$69 which
still has multiwave emissions.

The twisted magnetosphere model of magnetars (Thompson et al.
2002) showed that the effective magnetic field will be larger for
a larger twist. If the magnetosphere of PSR B0540$-$69 is twisted
by a glitch, then it will also result in a larger spin-down rate.
However, the twisted magnetosphere will relax back to the pure
magnetic dipole case in several years (Beloborodov 2009). During
this process, the neutron star X-ray luminosity, spin-down rate
will both decrease with time. For PSR B0540$-$69, its high
spin-down state has lasted more than 3 years (Marshall et al.
2015). This is inconsistent with the expectation of the twisted
magnetosphere model.

There are also external models for the braking index or
intermittent pulsar spin-down behavior, e.g., the fallback disk
model (Liu et al. 2014 and references therein; Li et al. 2006).
However, these external models are hard to verify or falsify.
Furthermore, accretion will halt the magnetospheric activities. In
the presence of accretion, it may be difficult to reconcile with
the radio emissions in PSR B0540$-$69 and in other pulsars with
braking index measured.

Observations of the timing behavior and pulse profile of some
pulsars indicate that the $\dot{\nu}$ modulation and the
pulse-shape variation are correlated, e.g., PSR B0910$+$16 (Perera
et al. 2015a) and PSR B1859$+$07 (Perera et al. 2015b). The
connection indicates that both these phenomena are magnetosphereic
origin (Lyne et al. 2010). Theory of a variable particle density
in the magnetosphere is successfully applied to explain the
spin-down behavior and emission property of the intermittent
pulsar (Kramer et al. 2006; Li et al. 2014). The mode changing and
nulling pulsar may be understood similarity because of the
detection of variation in spin-down rate of PSR J1717$-$4054
(Young et al. 2015) and a weak emission state in addition to its
bright and nulling states of PSRs J1853+0505 and J1107-5907 (Young
et al. 2014). For PSR B0540$-$69, the increase of particle density
in the magnetosphere will change the spin-down rate, as well as
the pulse profile. The radio giant pulse of PSR B 0540$-$69
(Johnston et al. 2004) may be caused by a larger out-flowing
particle density. Besides, different emission models (core, cone
and patch) are applied to explain the variable mean pulse profiles
(Lyne \& Manchester 1988). And, it is predicted that the
corotation of magnetosphere with pulsar may also affects the
emission properties (Wang et al. 2012b). Hence, the nonuniform
distribution of particles between these core and conal componnets
will change the pulse shape also. For PSR B0540$-$69, the pulse
profile has a broad double peak which can be described with two
Gaussians with a phase separation of $20\%$ (de Plaa et al.
1993).We could emphasize that: (i) if the particles distribute
uniform, the increase in out-flowing particle density may result
in the increase of pulse intensity, the ratio of these two
component keeps constant; (ii) if the particles distribute
nonuniform, both the pulse intensity and the ratio will change;
(iii) the coherent manner of radiated particles may affect the
pulse shape as well. The pulse profile and total flux in the high
spin-down state are needed to compare with them in the low
spin-down state.

\section{Conclusions}

The pulsar wind model is applied to explain the variable timing
behavior of PSR B0540$-$69 and PSR J1846$-$0258. Both the
persistent state braking index and the variable spin-down rate of
PSR B0540$-$69 are understandable. A larger particle density will
result in an increase in the spin down rate and predicts a smaller
braking index. And an increasing particle density will lead to a
lower braking index. For PSR B0540$-$69, in the vacuum gap model,
corresponding to the $36\%$ increase in the spin-down rate, the
relative increase in particle density is $88\%$. And the braking
index decreases to $1.79$. The same conclusion is obtained for the
different acceleration models. Since it has both a variable
spin-down rate and persistent state braking index measured, PSR
B0540$-$69 is very powerful in constraining different pulsar
spin-down mechanisms. Future observation of braking index in the
new spin-down state will provide further test on different
spin-down models and different particle acceleration models in the
wind braking scenario. For PSR J1846$-$0258, the variable timing
behavior of a net decreasing in spin down frequency
($\Delta{\nu}\approx -10^{-4} \rm \, Hz$) can be understood
similarly. And a changing rate of particle density
$\dot{\kappa}=1.68 \times 10^{-9} \rm \, s^{-1}$ will result in
the lower barking index $2.19$.

\section*{Acknowledgments}
The authors would like to thank R.X.Xu for discussions. H.Tong is
supported West Light Foundation of CAS (LHXZ201201), 973 Program
(2015CB857100) and Qing Cu Hui of CAS.

\label{lastpage}


\begin{thebibliography}{99}

\bibitem{Archibald2015a}
Archibald R. F., Kaspi V. M., Ng C. Y., et al., 2015a,  ApJ, 800,
33

\bibitem{Archibald2015b}
Archibald R. F., Kaspi V. M., Beardmore A. P., et al., 2015b,
arXiv:1506.06104

\bibitem{Becker2009}
Becker W., 2009, Neutron stars and pulsars, ASSL, 357, 91

\bibitem{Beloborodov2009}
Beloborodov A. M., 2009, ApJ, 703, 1044

\bibitem{Beskin2007}
Beskin V. S., \& Nokhrina E. E., 2007, Ap\&SS, 308, 569

\bibitem{Contopoulos2006} %spin-down%
Contopoulos I., \& Spitkovsky A., 2006, ApJ, 643, 1139

\bibitem{Cusumano2003}
Cusumano G., Massaro E., \& Mineo T., 2003, A\&A, 402, 647

\bibitem{}
de Plaa J., Kuiper L., Hermsen W., 2003, A\&A, 400, 1013



\bibitem{Espinoza2011} %PSR J 1734-3333%
Espinoza C. M., Lyne A. G., Kramer M., et al., 2011, ApJ, 741, L13

\bibitem{Ferdman2015}
Ferdamn R. D., Archibald R. F., \& Kaspi, V. M. 2015,
arXiv:1506.00182

\bibitem{Gavriil2008}
Gavriil F. P., Gonzalez M. E., Gotthelf E. V., et al., 2008,
Science, 319, 1802

\bibitem{Goldreich1969} %\rho_{GJ}%
Goldreich P., \& Julian W. H., 1969, ApJ, 157, 869

\bibitem{Harding1999}
Harding A. K., Contopoulos I., \& Kazanas D., 1999, ApJ, 525, L125

\bibitem{}
Johnston S., Romani R. W., Marshall F. E., et al., 2004, MNRAS,
355, 31

\bibitem{Kou2015}
Kou F. F. \& Tong H., 2015, MNRAS, 450, 1990

\bibitem{Kramer2006} %Intermittent pulsar%
Kramer M., Lyne A. G., OBrien J. T., et al., 2006, Science, 312,
549

\bibitem{Li2012b}
Li J., Spitkovsky A., \& Tchekhovskoy A., 2012, ApJL, 746, L24

\bibitem{Li2014} %Intermittent pulsar%
Li L., Tong H., Yan W. M., et al., 2014, ApJ, 788, 16

\bibitem{Li2006}
Li X. D., 2006, ApJ, 646, L139

\bibitem{Liu2014}%fall back disk%
Liu X. W., Xu R. X., Qiao G. J., et al., 2014, RAA, 14, 85

\bibitem{Livingstone2005}
Livingstone M. A., Kaspi V. M., \& Gavriil F. P., 2005, ApJ, 633,
1095

\bibitem{Livingstone2006}
Livingstone M. A., Kaspi V. M., Gotthelf E. V., et al., 2006, ApJ,
647, 1286

\bibitem{Livingstone2010}
Livingstone M. A., Kaspi V. M., \& Gotthelf E. V., 2010, ApJ, 710,
1710

\bibitem{Livingstone2011}
Livingstone M. A., Ng C.-Y., Kaspi V. M., et al., 2011, ApJ, 730,
66

\bibitem{}
Lyne A. G., Manchester R. N., 1988, MNRAS, 234, 477

\bibitem{}
Lyne A., Hobbs G., Kramer M., Stairs I., Stappers B., 2010,
Science, 329, 408

\bibitem{Lyne 2013} %Crab inclination angle%
Lyne A. G., Smith F. G., Weltevrede P., et al., 2013, Science,
342, 598

\bibitem{Lyne 2014} %Crab observation%
Lyne A. G., Jordan C. A., Smith F. G., et al., 2015, MNRAS, 446,
857

\bibitem{Marshall2015}
Marshall F. E., Guillemot L., Harding A. K., et al., 2015, ApJ,
807, L27

\bibitem{}
Perera B. B. P., Stappers B. W., Weltevrede P., et al., 2015a,
446, 1380

\bibitem{}
Perera B. B. P., Stappers B. W., Weltevrede P., et al., 2015b,
arXiv: 151004484

\bibitem{Philippov2014}
Philippov A., Tchekhovskoy, A., \& Li J. G., 2014, MNRAS, 441,
1879

\bibitem{Ruderman1975}
Ruderman M. A., \& Sutherland P. G., 1975, ApJ, 196, 51

\bibitem{TLK (2002)}
Thompson C., Lyutikov M., \& Kulkarni S. R., 2002, ApJ, 574, 332

\bibitem{Tong2015}
Tong H., 2015, arXiv:1506.04605

\bibitem{Tong2013}
Tong H., Xu R. X., Song L. M., et al., 2013, ApJ, 768,144

\bibitem{Wang 2012} %Crab observation%
Wang J., Wang N., Tong H., et al., 2012, Astrophys.Space Sci.,
340, 307

\bibitem{}
Wang P. F., Wang C., \& Han J. L., 2012b, MNRAS, 423, 2464

\bibitem{Wu2003} %wind model%
Wu F., Xu R. X., \& Gil J., 2003, A\&A, 409, 641

\bibitem{Xu2001}
Xu R. X., \& Qiao G. J., 2001, ApJ, 561, L85

\bibitem{Young2013}
Young N. J., Stappers B. W., Lyne A. G., et al., 2013, MNRAS, 429,
2569



\bibitem{} Young N. Y., Weltevrede P., Stappers B. W., et
al., 2014, 442, 2519

\bibitem{}
Young N. Y., Weltevrede P., Stappers B. W., et al., 2015, 449,
1495

\bibitem{Yue2007}
Yue Y. L., Xu R. X., \& Zhu W. W., 2007, Advance in Space
Research, 40, 1491

\bibitem{Zhang2000}
Zhang L., \& Cheng, K. S., 2000, A\&A 363, 575

\bibitem{zhang2001}
Zhang W., Marshall F. E., Gotthelf E. V., et al., 2001, ApJ, 554,
L177

\bibitem{Zhou2014}
Zhou E. P., Lu J. G., Tong H., et al., 2014, MNRAS, 443, 2705



\end{thebibliography}
\end{document}